\documentclass[review]{elsarticle}
\usepackage{lineno,hyperref}
\journal{Journal of \LaTeX\ Templates}

\usepackage{amsfonts}
\begin{document}

\begin{frontmatter}

%\corref{mycorrespondingauthor}

\title{New hierarchy of multiple soliton  solutions of the (2+1)-dimensional Sawada-Kotera equation}
\baselineskip 6mm
\author{Ruoxia Yao\fnref{myfootnote} }
\fntext[myfootnote]{Corresponding author: rxyao2@hotmail.com; rxyao@snnu.edu.cn}
\address{School of Computer Science, Shaanxi Normal University, Xi'an, 710119, China}
\author{Yan Li \fnref{Li}}
\fntext[Li]{Email: yanli@snnu.edu.cn}
\address{School of Computer Science, Shaanxi Normal University, Xi'an, 710119, China}
\author{Senyue Lou\fnref{lou}}
\fntext[lou]{Email: lousenyue@nbu.edu.cn}
\address{ School of Physical Science and Technology,  Ningbo University, Ningbo, 315211, China}

\begin{abstract}
A new transformation $u=4 ({\rm ln}f)_x$ that can formulate a quintic linear equation and a pair of Hirota's bilinear equations for the (2+1)-dimensional Sawada-Kotera (2DSK) equation is reported firstly, which enables one to obtain a new hierarchy of multiple soliton solutions of the 2DSK equation. It tells a crucial fact that a nonlinear partial differential equation could possess two hierarchies of multiple soliton solutions and the 2DSK equation is the first and only one found in this paper. The quintic linear equation is solved by a pair of Hirota's bilinear equations,  of which one is the (2+1)-dimensional bilinear SK equation obtained by $u=2 ({\rm ln}f)_{x}$, and the other is the bilinear KdV equation. The (1+1)-dimensional SK equation does not possess this property. As another example, a (3+1)-dimensional nonlinear partial differential equation possessing a pair of Hirota's bilinear equations, however only bearing one hierarchy of multiple soliton solutions is studied.
\end{abstract}
\end{frontmatter}

\baselineskip 6mm
\section{Introduction}
Solitons  of nonlinear partial differential equations play important roles in both fundamental theory and applications of  nonlinear science. As shown in \cite{Brazhnyi2005}, the soliton and solitary wave solutions that reflect a common nonlinear phenomenon in nature provide  physical information and more insights into the physical and mathematical aspects of the problem thus leading to further applications. In the last two decades, more and more researchers have taken their attentions to the study of solitary wave and soliton solutions of completely integrable  nonlinear evolution equations. Up to now, we know that an integrable nonlinear partial differential equation (PDE) usually possesses one hierarchy of multiple soliton  solutions, however we find that the  2DSK equation possesses two hierarchies of multiple soliton  solutions. This very special property is novel and first proposed in this paper.

The 2DSK equation
\begin{equation}\label{SK}
v_{t}+v_{5x}+15v_xv_{xx}+15vv_{3x}+45v^2v_x+5v_{xxy}+15vv_y+15v_{x}\int v_y {\rm d}x-5\int v_{yy} {\rm d}x=0,
\end{equation}
where $v=v(x, y, t)$, was first proposed by Konopelchenko and Dubrovsky \cite{KD1984}, and has been regarded as a (2+1)-dimensional integrable generalization of the (1+1)-dimensional  Sawada-Kotera equation \cite{Sawada1974PTP}. Now it is known also as a member  of the so-called CKP hierarchy \cite{Dubrovsk2002}. The 2DSK hierarchy with almost the same form and only having different coefficients of several terms of Eq. (\ref{SK}) are studied extensively, such that abundant results have been reported in the Refs. \cite{Geng1989, Lou1994, Dubrovsk2002, Cao2008, WazwazX2010, LvX2010, ShiYQ2012, LvX2018, Jia2017} and therein. Recently, many new important types of solutions such as lump solutions, molecule soliton solutions and resonant soliton solutions are reported constantly in Refs. \cite{MaWX2017, LLHuang2017, LiX2017, AnHL2019, KuoCK2019}.
\section{New transformation and a pair of Hirota's bilinear equations of the 2DSK equation}
Eq. (\ref{SK}) reduces to
\begin{equation}\label{SK1}
u_{xt}+u_{6x}+5u_{3xy}-5u_{yy}+15u_{xx}u_{3x}+15u_xu_{4x}+15u_{x}u_{xy}+15u_{xx}u_{y}+45u_x^2u_{xx}=0,
\end{equation}
after introducing a potential variable transformation $v=u_x, u=u(x, y, t),$   and $u_{xt}={\frac {\partial ^{2}}{\partial x\partial t}}u \left( x,y,t \right)$. Eq. (\ref{SK1}) is the two (space)-dimensional extension of
the famous KP equation first reported by Kadomtsev and Petviashvili \cite{Kadomtsev1970} . In this paper, we mainly consider the reduced Eq. (\ref{SK1}). The 2DSK equation usually can be written in Hirota's bilinear form by expressing solutions in terms of a $\tau$-function. If  $u=2 ({\rm ln}\tau)_{x}$
then $\tau(x, y, t)$ satisfies the Hirota's bilinear equation
\begin{equation}\label{Obin}
 (D_{x}^6+D_{x}D_t-5D_{x}^3D_y+5D_y^2)\tau \cdot \tau=0,
\end{equation}
where the Hirota's $D$-operator is defined in Ref. \cite{Hirota2004}. In the following, we will show that we obtain different bilinear expressions of Eq. (\ref{SK1}).

In Refs. \cite{JMPHt1987-1, JMPHt1987-2, JMPHt1988, JMPHt1989}, Hietarinta provided a complete classification for KdV-type,
mKdV-type, sine-Gordon-type and complex bilinear equations passing Hirota's three-soliton condition. In Ref. \cite{ChengJMP1992}, Cheng studied some decompositions including (1+1)-dimensional KdV+$5th$-order KdV and Ito+KdV equations. The results presented in this paper have not yet been reported therein or elsewhere.

Starting with a dependent variable transformation
\begin{equation}\label{tr1}
  u=p({\rm ln}f)_{x}
\end{equation}
 where $f=f(x,y,t)$ and combining with the assumption
\begin{equation}\label{sf1}
  f\equiv f_1=1+{\rm exp}(\xi_1), \hskip 5mm \xi_1=k_1 x+l_1 y+w_1 t,
\end{equation}
instead of using the Hirota's bilinear method, we get (see Ref. \cite{CTPYao2004} for more details)
\begin{eqnarray}
  && p =  2,\,\, w_1 = -(k_1^6+5 k_1^3 l_1-5l_1^2)/k_1,\label{p1}\\
  && p =  4,\,\, l_1 =-k_1^3,\,\, w_1 = 9 k_1^5,\label{p2}
\end{eqnarray}
 and then a quintic linear equation for Eq. (\ref{SK1}) later corresponding to (\ref{p2}). The advantage is that the direct method could avoid seeking for bilinear form at the beginning of constructing multiple soliton solutions. We prove that even a nonlinear PDE does not possess the so-called Hirota's bilinear form apparently \cite{HirotaPRL1971, HirotaJPSJ1976, Hirota2004}, maybe we could obtain a new hierarchy of related soliton solutions.

In this paper, we adopt the determined dependent variable transformation
\begin{equation}\label{trs2}
 u=4 ({\rm ln}f)_{x}
\end{equation}
instead of $u=2 ({\rm ln}f)_{x}$ to find soliton solutions  first of Eq. (\ref{SK1}), and then use $v \rightarrow u_x$, namely, $v=4 ({\rm ln}f)_{xx}$ to obtain the multiple soliton  solutions of Eq. (\ref{SK}). Substituting Eq. (\ref{trs2}) into Eq. (\ref{SK1}) yields the following quintic linear equation
\begin{eqnarray}
% \nonumber to remove numbering (before each equation)
  P \equiv && f^4(f_{7x}+5f_{4xy}+f_{2xt}-5f_{x2y})+f^3\left(25f_{4x}f_{3x}+39f_{5x}f_{2x}-5f_{4x}f_{y}-f_{2x}f_{t}-7f_{6x}f{x}\right.\nonumber\\ &&\left.-20f_{3xy}f_{x}+40f_{3x}f_{xy}+5f_{x}f_{2y}+30f_{2x}f_{2xy}+10f_{xy}f_{y}-2f_{xt}f_{x}\right)\nonumber\\
  && +2f^2\left(-10f_{3x}f_{x}f_{y}+f_{x}^2f_{t}-50f_{3x}^2f_{x}-135f_{4x}f_{x}f_{2x}-9f_{5x}f_{x}^2-90f_{2x}f_{x}f_{xy}+75f_{3x}f_{2x}^2\right.\nonumber\\
  && \left.-5f_{x}f_{y}^2-15f_{2x}^2f_{y}\right)+2f(90f_{2x}f_{x}^2f_{y}+270f_{3x}f_{x}^2f_{2x}+60f_{x}^3f_{xy}-225f_{2x}^3f_{x}+105f_{4x}f_{x}^3)\nonumber\\
   &&-40(12f_{3x}f_{x}^4+9f_{2x}^2f_{x}^3-3f_{x}^4f_{y})=0.\label{5LForm}
  \end{eqnarray}
Therefor, if $f=f(x, y, t)$ solves Eq. (\ref{5LForm}), $u=4 ({\rm ln}f)_{x}$ solves Eq. (\ref{SK1}), and $v$ obtained by computing once derivative of the obtained solution $u$ with respect to $x$ solves Eq. (\ref{SK}).

\bf Theorem 1 \rm \em If $f=f(x,y,t)$ solves  a couple of bilinear form equations
\begin{eqnarray}
(D_{x}^6+D_{x}D_t+5D_{x}^3D_y-5D_y^2)  f\cdot f =0,\label{biForm1}\\
 (D_{x}^{4}+D_xD_y)  f\cdot f =0,\label{biForm1-1}
\end{eqnarray}
where Eq. (\ref{biForm1}) is of the standard  Hirota's  bilinear 2DSK equation, and  Eq. (\ref{biForm1-1}) is of the standard Hirota's bilinear KdV equation, then $u=4 ({\rm ln}f)_{x}$ solves Eq. (\ref{SK1}) with $f$ also solved by  the quintic linear equation (\ref{5LForm}).
\rm \\

{\bf Proof}  From the definition of the Hirota's bilinear derivative, the quintic linear equation (\ref{5LForm}) can be written  as the following form
  \begin{eqnarray}
  P &\equiv &2(\Delta_{\rm sk})_x/f^2 -4 \left[\Delta_{\rm sk}f_x -15\Delta_{\rm kdv}\left(f_{xx}+f_{3x}\right)\right]/f^3 \nonumber\\
 && -60\Delta_{\rm kdv} \left[\left(f_x^2+5f_{xx} f_x\right)/f^4 -4 f_x^3/f^5\right],\label {EquivBiForms0}
  \end{eqnarray}
where
$\Delta_{\rm kdv} \equiv (D_x^4  +D_{x}D_y)  f\cdot f ,\quad
\Delta_{\rm sk} \equiv  ( D_x^6  +D_{x}D_t +5 D_{x}^3D_y -5 D_y^2)  f\cdot f.$
 It shows that if  $\Delta_{\rm sk}=0$ and $\Delta_{\rm kdv}=0$, then $P=0$. Obviously, Eq. (\ref{biForm1}), namely  $\Delta_{\rm sk}$,  is the standard  Hirota's  bilinear 2DSK equation, and  Eq. (\ref{biForm1-1}),  $\Delta_{\rm kdv}$, is  the  Hirota's bilinear KdV equation.

To show that $f=f(x,y,t)$ solving Eqs. (\ref{biForm1}) and  (\ref{biForm1-1}) is a solution of Eq. (\ref{5LForm}), one should prove the  consistency of Eqs. (\ref{biForm1}) and (\ref{biForm1-1}), say, of which  the solution set is not empty.
For simplicity and without loss of generality,  setting $f={\rm exp}({\tilde v}), {\tilde v}={\tilde v}(x,y,t)$ in Eqs. (\ref{biForm1}), (\ref{biForm1-1}) and solving ${\tilde v}_{xt},\, {\tilde v}_{yx}$ respectively yields
\begin{equation}\label{vxtvxy}
  {\tilde v}_{xt}=120{\tilde v}_{2x}^3-5{\tilde v}_{3xy}+5{\tilde v}_{2y}-{\tilde v}_{6x},\quad {\tilde v}_{xy}=-{\tilde v}_{4x}-6{\tilde v}_{2x}^2.
\end{equation}
 Next, one should prove ${\tilde v}_{xt,xy}={\tilde v}_{xy,xt}.$ Easily, we get
 \begin{equation}\label{vxytx}
 \begin{array}{l}
   {\tilde v}_{xy,xt}=-{\tilde v}_{5xt}-12{\tilde v}_{3x}{\tilde v}_{2xt}-12{\tilde v}_{2x}{\tilde v}_{3xt},\\
    {\tilde v}_{xt,xy}=720{\tilde v}_{2x}{\tilde v}_{2xy}{\tilde v}_{3x}+360{\tilde v}_{2x}{\tilde v}_{3xy}-5{\tilde v}_{4x2y}+5v_{x3y}-{\tilde v}_{7xy}.
   \end{array}
 \end{equation}
  Computing ${\tilde v}_{xy,xt}-{\tilde v}_{xt,xy}$ and substituting ${\tilde v}_{xt}$ and ${\tilde v}_{xy}$ with the forms (\ref{vxtvxy}) into the obtained expression, one can obtain ${\tilde v}_{xy,xt}-{\tilde v}_{xt,xy}\equiv 0. \quad\quad \Box$

Theorem 1 shows that the obtained quintic linear equation is solved by the couple of Hirota's bilinear equations which usually serves as a role to construct the new hierarchy of soliton solutions. It is well known that Eq. (\ref{biForm1})  ensures the $N$-soliton solutions of the 2DSK equation starting from $u=2 ({\rm ln}f)_x$, and Eq. (\ref{biForm1-1}) ensures the $N$-soliton solutions of the KdV equation. Therefore, an explicit connection between the KdV equation and the 2DSK equation is established.
\section{Two hierarchies of multiple solition solutions of the 2DSK equation}
To construct multiple soliton  solutions of the 2DSK Eq. (\ref{SK1}), usually one can start from Eq. (\ref{trs2}) with \cite{Hirota2004}
\begin{eqnarray}
% \nonumber to remove numbering (before each equation)
  f_1&=&1+{\rm exp}(\xi_1), \label{skf1}\\
  f_2&=&1+{\rm exp}(\xi_1)+{\rm exp}(\xi_2)+h_{1,2}{\rm exp}(\xi_1+\xi_2),\label{skf2}\\
  f_3&=&1+{\rm exp}(\xi_1)+{\rm exp}(\xi_2)+{\rm exp}(\xi_3)\nonumber\\
  &&+h_{1,2}{\rm exp}(\xi_1+\xi_2)+h_{1,3}{\rm exp}(\xi_1+\xi_3)+h_{2,3}{\rm exp}(\xi_2+\xi_3)\nonumber\\
  &&+h_{1,2}h_{1,3}h_{2,3}{\rm exp}(\xi_1+\xi_2+\xi_3),\label{skf3}\\
&&\ldots\ldots, \nonumber\\
 f_n&=&\sum_{\mu\in{0, 1}}{\rm exp}\left(\sum_{i=1}^{n}\mu_i\xi_i+\sum\mu_i\mu_jH_{ij}\right), \label{skfn}
\end{eqnarray}
where $\xi_i=\xi_i(x,y,t)=k_i x+l_i y+w_i t, (i=1, \ldots, n)$, and $k_i, l_i, w_i$ are arbitrary constants for all $i$. It is actually a simplified direct Hirota method.

Usually, researchers proceed with the studies such as constructing multiple soliton  solutions by taking  $p=2$ in  Eq. (\ref{tr1}). The transformation $u=2 ({\rm ln}f)_{x} $  enables one to obtain a Hirota's bilinear form, and then soliton solutions. It is deemed that one could not obtain high-order multiple soliton  solutions if $u=4({\rm ln}f)_{x}$ does not make the 2DSK to transform into a Hirota's bilinear form directly. It is crucial that we do obtain a new hierarchy of multiple soliton  solutions of the 2DSK equation, which have not been given in previous references. The 2DSK equation (\ref{SK1}) is the first and only one that bearing two hierarchies of multiple soliton solutions.

For completeness, two one-soliton solutions of Eq. (\ref{SK1}) are listed bellow
\begin{equation}\label{SS1}
\begin{array}{l}
  u_1=4({\rm ln}f_1)_{x} \hskip 2mm {\rm with} \hskip 2mm f_1=1+{\rm exp}(k_1 x-k_1^3 y+9 k_1^5 t),\\
{\tilde u}_1 =2({\rm ln}{\tilde f}_1)_{x} \hskip 2mm {\rm with} \hskip 2mm {\tilde f}_1=1+{\rm exp}\left(k_1 x+l_1  y+\frac{5 l_1^2-k_1^6+5k_1^3 l_1}{k_1} t\right),
\end{array}
\end{equation}
where $k_1\neq 0, l_1\neq 0$ are free parameters. They can be written as the following forms \cite{JMP2018Lou}.
\begin{equation}\label{SS1-1}
\begin{array}{l}
u_1=4{\rm ln}\left({\rm cosh}(\frac{k_1 }{2}x-\frac{k_1^3}{2}y+\frac{9 k_1^5}{2}t)\right)_x, \\
{\tilde u}_1=2{\rm ln}\left({\rm cosh}(\frac{p_1+q_1 }{2}x-\frac{p_1^3+q_1^3}{2}y+\frac{9 (p_1^5+q_1^5)}{2}t)\right)_x,
\end{array}
\end{equation}
where $p_1, q_1$ are arbitrary constants.
\subsection{Two-soliton solutions of the 2DSK equation}
As for two-soliton solutions, starting from $u=4 ({\rm ln}f_{2\rm new})_{x}$ of Eq. (\ref{SK1}) we obtain a new two-soliton solution with  $f_{2\rm new}$ having the  form
\begin{equation}\label{2S-new}
f_{2\rm new}= a_{1,2} {\rm cosh}(\frac{\xi_1}{2}+\frac{\xi_2}{2})+b_{1,2} {\rm cosh}(\frac{\xi_1}{2}-\frac{\xi_2}{2}), \quad \xi_i=k_i x+l_i y+w_i t, \end{equation} and
$l_i=-k_i^3,\, w_i=9 k_i^5,\,(i=1, 2), \,a_{1,2}=k_1-k_2,\,b_{1,2}=k_1+k_2, \,k_1\neq \pm k_2$.

As a comparison, we give the  two-soliton solution in terms of \rm{\it cosh} function  obtained  by $u=2 ({\rm ln}f_{2\rm old})_{x}$
with $f_{2\rm old}$ having the same form with that of Eq. (\ref{2S-new}), whereas
\begin{equation}\label{2S-Old}
\begin{array}{l}
k_i=p_i+q_i,\quad l_i=-(p_i^3+q_i^3),\quad w_i=9(p_i^5+q_i^5),\quad (i=1, 2), \\
a_{1,2}=\sqrt{(q_1-q_2)(q_1-p_2)(p_1-q_2)(p_1-p_2)},\\
b_{1,2}=\sqrt{(q_1+q_2)(q_1+p_2)(p_1+q_2)(p_1+p_2)},
\end{array}
\end{equation}
where, and from now on, $p_i, q_i,\,(p_i\neq \pm p_j, \, q_i\neq \pm q_j,\,p_i\neq \pm q_j) $ are constants for all $i$.
\subsection{Three-soliton solutions of the 2DSK equation}
Starting from $u=4 ({\rm ln}f)_{x}$ we obtain a new three-soliton solution with
\begin{eqnarray}
% \nonumber to remove numbering (before each equation)
  f_{3\rm new}\!\!\!&=&\!\!\! K_0 {\rm cosh}(\frac{\xi_1}{2}+\frac{\xi_2}{2}+\frac{\xi_3}{2})+K_1 {\rm cosh}(\frac{\xi_1}{2}-\frac{\xi_2}{2}-\frac{\xi_3}{2})+ \nonumber\\
  \!\!\!&&\!\!\!K_2 {\rm cosh}(\frac{\xi_1}{2}-\frac{\xi_2}{2}+\frac{\xi_3}{2})+K_3 {\rm cosh}(\frac{\xi_1}{2}+\frac{\xi_2}{2}-\frac{\xi_3}{2}),\label{coshformNew3}
  \end{eqnarray}
where $\xi_i=k_i x-k_i^3 y+9 k_i^5 t, \,(i=1,\ldots, 3), $ and
\begin{equation}\label{K0123}
\begin{array}{l}
  K_0=a_{1,2}a_{1,3}a_{2,3},\quad K_1=b_{1,2}b_{1,3}a_{2,3},\quad K_2=b_{1,2}a_{1,3}b_{2,3},\quad K_3=a_{1,2}b_{1,3}b_{2,3},\\
a_{i,j}=k_i-k_j,\quad b_{i,j}=k_i+k_j,\quad k_i\neq \pm k_j,\quad (1\leq i<j\leq 3).
\end{array}
\end{equation}

For the known three-soliton solution in terms of $\rm{\it cosh}$ function, $f_{3\rm old}$ has the same form with that of Eq. (\ref{coshformNew3}),
where $$\xi_i=k_i x+l_i y+w_i t,\quad k_i=p_i+q_i,\quad l_i=-(p_i^3+q_i^3),\quad w_i=9(p_i^5+q_i^5),\quad (i=1,\ldots, 3),$$ and $K_i\, (i=0,\ldots, 3)$ is the same with (\ref{K0123}), however $a_{i,j},\,b_{i,j}$ have the following different forms
$$
\begin{array}{l}
% \nonumber to remove numbering (before each equation)
  a_{1,2}=\sqrt{(q_1-q_2)(q_1-p_2)(p_1-q_2)(p_1-p_2)},\quad b_{1,2}=\sqrt{(q_1+q_2)(q_1+p_2)(p_1+q_2)(p_1+p_2)}, \\
  a_{1,3}=\sqrt{(q_1-q_3)(q_1-p_3)(p_1-q_3)(p_1-p_3)},\quad b_{1,3}=\sqrt{(q_1+q_3)(q_1+p_3)(p_1+q_3)(p_1+p_3)},\\
  a_{2,3}=\sqrt{(q_3-q_2)(q_3-p_2)(p_3-q_2)(p_3-p_2)},\quad b_{2,3}=\sqrt{(q_2+q_3)(p_2+q_3)(q_2+p_3)(p_2+p_3)}.
\end{array}
$$
\subsection{$N$-soliton solutions of the 2DSK equation}
Starting from $u=4({\rm ln} f_{N\!{\rm new}})_{x}$ we obtain an $N$-soliton solution with
\begin{eqnarray}
 f_{N\!{\rm new}}&=&\sum_{\nu}{K_\nu}{\rm cosh}\left( \sum_{i=1}^{N}\frac{\nu_i\xi_i}{2}\right),\quad  K_\nu=\prod_{i<j}a_{i,j},\label{Nnew}
 \end{eqnarray}
 where $\xi_i=k_i x-k_i^3 y+9 k_i^5 t,\,(i=1,\ldots, N), \, a_{i,j}=k_i-\nu_i\nu_jk_j,\, (1\leq i<j\leq N),$ and  the summation of $\nu=\{\nu_1, \nu_2, \ldots, \nu_N \}$ are did for all non-dual permutations of $\nu_i=1, -1, (i=1, \ldots, N),\, (\nu $ $\nu'$ are dual if $\nu=-\nu').$

 The already existing $N$-soliton solution that we can write it in terms of $\rm{\it cosh}$ function, $ f_{N\!{\rm old}}$ is the same with that of ({\ref{Nnew}), and  $\xi_i=k_i x+l_i y+w_i t,\,(i=1,\ldots, N), \,k_i=p_i+q_i,\, l_i=-(p_i^3+q_i^3),\, w_i=9(p_i^5+q_i^5)$ with
 $$ a_{i,j}^2=(q_i-\nu_i\nu_jq_j)(q_i-\nu_i\nu_jp_j)(p_i-\nu_i\nu_jq_j)(p_i-\nu_i\nu_jp_j),$$
and  the summation of $\nu=\{\nu_1, \nu_2, \ldots, \nu_N \}$ same as before.
\section{Concluding remarks}
In short, this paper is devoted to exploring new auxiliary transformation to derive a quintic linear equations that can be solved by a pair of Hirota's bilinear forms used for constructing new hierarchy of multiple soliton solutions of the 2DSK equation. Also, an explicit connection between the KdV and the 2DSK equations is established. Up to now, the 2DSK is the only nonlinear PDE reported in this paper that possesses two hierarchies of multiple soliton solutions. The new transformation  has not been given more attention or has been missed by researchers before maybe because it can not provide Hirota's bilinear form directly by using the classical Hirota's bilinear method. We implement this newly found simple but novel transformation to find new hierarchy of multiple soliton solutions of the 2DSK equation. The famous 2DSK equation is investigated successfully, and for further we hope to obtain a category of nonlinear PDEs that also possess multiple soliton solutions obtained by two or three Hirota's bilinear equations.
 Actually, there indeed exists such nonlinear partial differential equation. As an example, we consider a (3+1)-dimensional nonlinear PDE with the form
\begin{equation}\label{3+1Eq}
\begin{array}{l}
 v_{2x}v_{yz}+2 v_{2x}v_{3xz}+ 12 v_{xx}^2v_{xz}+ v_{xy}v_{xt}-120 v_{xy}v_{4x}v_{2x}-4 v_{xy}v_{6x}-240 v_{xy}v_{2x}^3\\
-10 v_{xy}v_{3xy}-60 v_{xy}v_{2x}+ 5 v_{xy}v_{2y}=0.
\end{array}\end{equation}
Likewise, substituting $v={\rm ln}(f),\,f=f(x, y, z, t)$ into the Eq. (\ref{skf3}), we obtain the following couple of bilinear equations
\begin{equation}\label{3bf}
(D_y D_z+2D_x^3D_z)f\cdot f =0,\quad (D_x D_t-4 D_x^6-10 D_x^3D_y+5 D_y^2)f\cdot f =0.
\end{equation}
The consistency of them is discussed bellow.
From the bilinear equations in (\ref{3bf}), we get
\begin{equation}\label{3bf1}
\begin{array}{l}
-8f_{6x}f+48f_{5x}f_x-120f_{4x}f_{2x}+80f_{3x}-20f_{3xy}f+60f_{2xy}f_x-60f_{xy}f_{2x}\\+20f_yf_{3x}+2f_{xt}f-2f_xf_t
-2f_xf_t-10f_{2y}f+10f_{y}^2=0,\\
4 f_{3xz}f-12 f_{2xz}f_x+12 f_{xz}f_{2x}-4 f_{3x}f_z+2f_{yz}f-2 f_y f_z=0.
\end{array}
\end{equation}
Same as Theorem 1, we can get
\begin{equation}\label{vxt}
  {\tilde v}_{xt}=240{\tilde v}_{2x}^3+10 {\tilde v}_{3xy}+120{\tilde v}_{4x}{\tilde v}_{2x}-5{\tilde v}_{2y}+4 {\tilde v}_{6x},\quad {\tilde v}_{yz}=-2 {\tilde v}_{3xz}-12 {\tilde v}_{2x}{\tilde v}_{xz}.
\end{equation}
 It is easy to prove that $({\tilde v}_{xt})_{yz}-({\tilde v}_{yz})_{xt}=0$ after simple computations, which completes the proof of the consistency. Thus, three-soliton solution of Eq. (\ref{3+1Eq}) by $v={\rm ln}(f)$ with $f$ being the same form given in Eq. (\ref{skf3}) is obtained, whereas,
$$
\begin{array}{l}
\xi_i=k_i x+p_i y+q_i z+w_i t,\quad p_i=-2 k_i^3,\quad w_i=-36 k_i^5,\quad  (i=1, \ldots, 3)\\
h_{i,j}=\frac{(q_i-q_j)(k_i-k_j)}{(q_i+q_j)(k_i+k_j)},\quad q_i\neq \pm q_j,\quad k_i\neq \pm k_j,\quad (1\leq i<j\leq 3).\end{array}
$$

\vskip 0.3cm \noindent {\Large{\bf Acknowledgements}}
\vskip 0.2cm
\noindent The work is supported by the National Natural Science Foundation of China (11471004, 11975131).

\bibliographystyle{elsarticle-num}
\section*{References}

\bibliography{yaorefs1}
\end{document}